 \documentclass[final,5p,times,twocolumn,authoryear]{elsarticle}

\usepackage{amsmath}
\usepackage{amssymb}
\usepackage{lipsum}
\usepackage{caption}
\journal{International Journal of Geometric Methods in Modern Physics}
\numberwithin{equation}{section}

\begin{document}

\begin{frontmatter}



\title{Degenerate Geometries as Matter-Free Physical Configurations \\
in General Relativity: Three Examples
}


\author[first]{Juri Dimaschko}
\ead{dimaschko@gmx.net}
\address{Technische Hochschule Lübeck, Mönkhofer Weg 239, 23562 Lübeck, Germany}
\begin{abstract}
We examine three degenerate spacetime configurations with wormhole topology, obtained via branching coordinate transformations of the Rindler, Minkowski, and Schwarzschild vacuum metrics. These configurations are, respectively, a Rindler wormhole with a planar throat, and the Klinkhamer and Schwarzschild–Klinkhamer wormholes with spherical throats. Within the framework of the Einstein–Palatini–Cartan formulation, we demonstrate that these degenerate configurations are matter-free. In this regard, they differ fundamentally from their non-degenerate wormhole counterparts in the thin-shell model, which require exotic matter. Nevertheless, the degenerate configurations considered here exhibit distinct physical manifestations, ranging from purely topological structures to genuine gravitational effects. Furthermore, in all three examples, we demonstrate the absence of a limiting transition from the non-degenerate spacetime to the matter-free degenerate configuration. This suggests that degenerate geometries constitute an independent sector of the configuration space of general relativity.
   
\end{abstract}



\begin{keyword}
%% keywords here, in the form: keyword \sep keyword, up to a maximum of 6 keywords
Einstein-Palatini-Cartan action \sep wormhole \sep degenerate metric \sep   Rindler metric \sep Klinkhamer metric  \sep Weyl curvature \sep thin shells



\end{keyword}

\end{frontmatter}

%\tableofcontents



\section{Introduction}

In general relativity, gravity is described by the geometry of spacetime, and the sources of the gravitational field are typically associated with the stress–energy tensor through Einstein's equations.

At the same time, the nonlinear nature of the theory permits the existence of self-consistent spacetime configurations that do not require matter in the conventional sense. In such cases, nontrivial geometric or physical effects may arise purely from the structure of spacetime itself. The aim of this work is to analyze such matter-free configurations in the context of \textit{degenerate} geometries.

One of the first examples of such configurations appears in the work of Einstein and Rosen [Einstein, 1935], where both the Rindler metric and the Einstein–Rosen bridge were analyzed. In these constructions, the presence of \textit{degenerate} hypersurfaces plays a significant role, allowing configurations that do not require explicit material sources.

Historically, however, the development of general relativity took place within the framework of Riemannian geometry, where the metric is assumed to be \textit{non-degenerate}. Fundamental results in this field impose significant restrictions on the existence of nontrivial configurations. These include Lichnerowicz's theorem on stationary vacuum solutions \citep{lichnerowicz}, Geroch's results on the impossibility of topological change without singularities or causality violation \citep{geroh}, and energy dominance conditions for traversable wormholes \citep{hochberg}. Crucially, all these results rely on the assumption of a \textit{non-degenerate} metric.

The possibility of systematically incorporating degenerate configurations into the variational formulation of gravity was proposed by Peres and Katanaev \citep{peres,katanaev} and further developed in Horowitz's approach \citep{horowitz}, based on the Einstein–Palatini–Cartan action and the tetrad formalism. In this approach, \textit{degenerate} and \textit{non-degenerate} configurations are considered on an equal footing, thereby expanding the configuration space of the theory without altering existing solutions in the non-degenerate sector.

In recent years, interest in \textit{degenerate} configurations has increased in connection with the work of Klinkhamer et al. \citep{klinkhamer1,klinkhamer2,klinkhamer3,wang}, who investigated the so-called defect wormholes. These solutions possess a nontrivial topology and contain \textit{degenerate} hypersurfaces that act as geometric defects. Their similarity to thin-shell models has led to active discussion and criticism \citep{baines,feng}, particularly with regard to the interpretation of the throat structure and the possible presence of matter on it.

Subsequent studies \citep{dimaschko2} have shown that \textit{degenerate} geometries form an independent sector of states in general relativity that does not require the presence of matter. This raises the question of how such matter-free states manifest themselves physically and how they should be described.

The goal of this paper is to address this question within the Einstein–Palatini–Cartan formalism. Unlike the standard metric formulation, this approach allows one to consider both \textit{degenerate} and \textit{non-degenerate} states within a unified framework without introducing additional assumptions.

The primary approach employed in this work is the method of topological dressing \citep{dimaschko}. Based on a branching coordinate transformation, this method transforms a known vacuum solution of the Einstein equations into a two-sheeted \textit{degenerate} wormhole geometry with a prescribed throat shape. In this respect, the method of topological dressing differs fundamentally from the thin-shell model, which operates only with \textit{non-degenerate} metrics.

In Section 2 of this paper, we provide a brief description of this method and compare it with the thin-shell model. Subsequently, we use the method of topological dressing to construct three \textit{degenerate} metrics describing the Rindler wormhole (Section 3), the Klinkhamer wormhole (Section 4), and the Schwarzschild-Klinkhamer wormhole (Section 5). In all three cases, it is demonstrated that the Ricci tensor vanishes throughout the entire spacetime - including the throat- which corresponds to the absence of matter in the standard sense.

Despite the absence of matter, the configurations under consideration exhibit distinct yet clearly defined manifestations: a purely topological structure (the Klinkhamer wormhole), gravitational acceleration in the absence of Weyl curvature (the Rindler wormhole), and a gravitational field with non-zero Weyl curvature (the Schwarzschild-Klinkhamer wormhole).

Furthermore, for each configuration, a regularized \textit{non-degenerate} metric is introduced; this metric depends on a parameter \(\epsilon\) and reduces to the corresponding (\textit{degenerate}) metric in the limit \(\epsilon=0\). An analysis of the action and the associated mass of matter reveals that the limit as \(\epsilon \rightarrow 0\).  does not coincide with the value obtained at \(\epsilon=0\) . Specifically, in the \textit{non-degenerate} case, a matter distribution emerges that is consistent with the thin-shell model, whereas in the strictly \textit{degenerate} state, matter is absent.

These results indicate that degenerate wormholes can be regarded as self-consistent configurations in which the topology and curvature of spacetime are not related to the presence of matter in the usual sense. At the same time, the \textit{degenerate} state does not coincide with the limit of smooth metrics and should therefore be treated as an independent element of the configuration space. This provides a natural distinction between matter-free \textit{degenerate} configurations and \textit{non-degenerate} configurations that require matter, such as those described within the thin-shell formalism. In this sense, \textit{degenerate} states following from the topological dressing method and \textit{non-degenerate} states following from the thin shell model should be viewed not as two alternative descriptions of the same object, but as two fundamentally different types of spacetime structure.

Throughout the article, a unit system is used in which the speed of light \(c\) and the gravitational constant \(G\) are equal to unity: \(c=G=1\).

\section{Two geometric frameworks: thin shells and topological dressing}

\textit{In this section, we briefly describe the topological dressing method \citep{dimaschko}, which constructs a two-sheeted wormhole geometry with a prescribed throat location \(\Sigma\) by means of a two-sheeted coordinate transformation \(x \mapsto x'\), for which the hypersurface \(\Sigma\) becomes a branching surface.}

To clarify the geometric meaning of this construction, we compare the topological dressing approach with the well-known thin-shell approach, which also generates two-sheeted wormhole geometries.

The thin-shell model entails two steps:

i) we take a known one-sheeted solution to the vacuum Einstein equations and excise identical regions - bounded by a closed surface \(\Sigma\) - from two copies of the same spacetime;

ii) subsequently, we glue the two sheets together along the boundary \(\Sigma\), that is, we identify points  on the boundary \(\Sigma\) in two distinct sheets that share identical coordinates.

The resulting two-sheeted geometry possesses the topology of a wormhole, as shown in Fig. 1. In this construction, the wormhole throat \(\Sigma\) arises as a junction surface between two glued spacetime sheets.

\begin{figure*}[t]
\centering
\captionsetup{skip=2pt}
\includegraphics[width=0.95\textwidth, trim=0 1.2cm 0 0, clip]{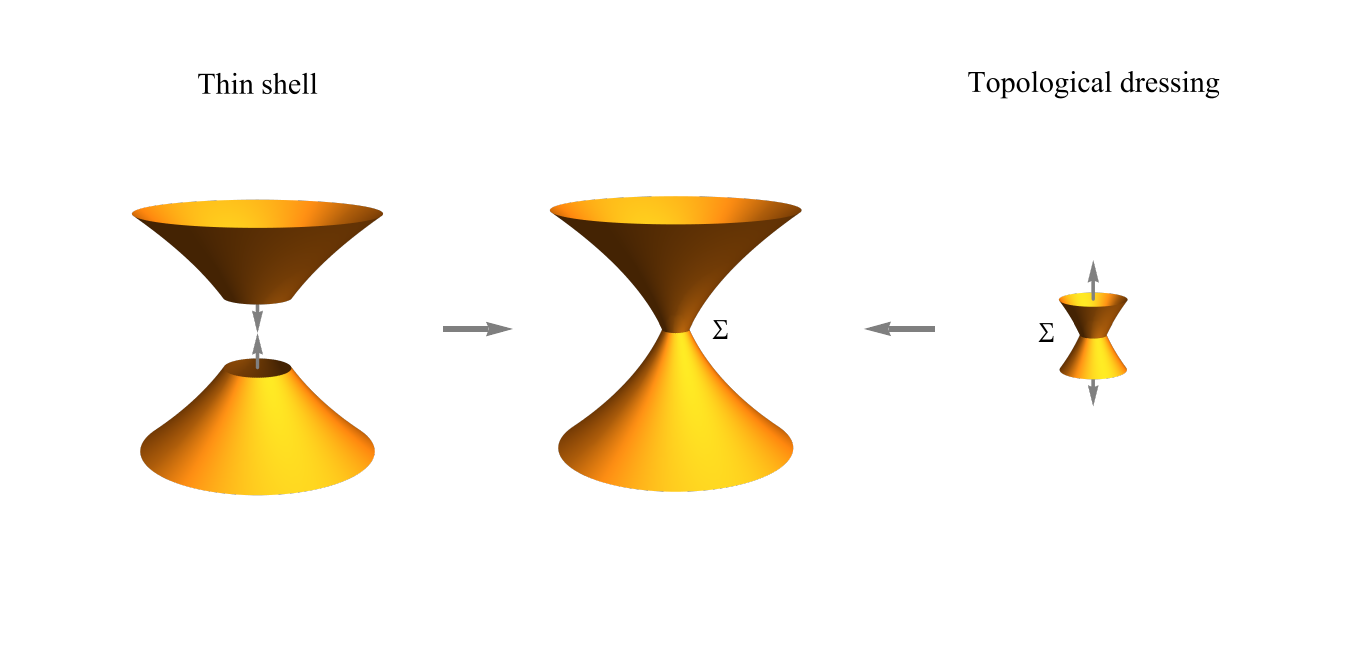}
\caption{Two geometric constructions of a two-sheeted wormhole configuration. In the thin-shell model, the wormhole is obtained by gluing together two spacetime regions along the throat \(\Sigma\), which acts as a junction surface. In the topological dressing method, the wormhole emerges through the branching of a single spacetime at the throat \(\Sigma\), which acts as a branching surface. The resulting configuration of the wormhole beyond the throat \(\Sigma\) is the same for both approaches. The embedding diagram is shown schematically and is inspired by the Schwarzschild–Klinkhamer wormhole geometry discussed in Section 5.}
\end{figure*}

An essential component of the thin-shell model is the matching of the metric across the boundary \(\Sigma\). To this end, Israel’s boundary conditions \citep{israel} are employed; these conditions determine the requisite surface stress-energy tensor \(S^i_j\). Typically, this results in a violation of the Null Energy Condition and the presence of exotic matter at the throat \citep{hochberg}.

Thus, the primary result of the thin-shell model is the surface stress-energy tensor \(S^i_j\) required to sustain an \textit{a priori} specified wormhole geometry.

The topological dressing method—just like the thin-shell model—involves two steps:

i) We take a known one-sheeted solution to the vacuum Einstein equations and select a region within this spacetime whose boundary is \(\Sigma\);

ii) We select a \textit{branching coordinate transformation} 
\begin{equation}
   f: x \mapsto \bar{x},
\end{equation}for which the boundary \(\Sigma\) is a branch surface, and the exterior of \(\Sigma\) is mapped into a two-sheeted geometry.

The branching transformation \(f\) is not a diffeomorphism, as it is not single-valued (it is two-valued). Its inverse transformation, \(f^{-1}\), is two-sheeted; its Jacobian, \(J = \det \left\lVert \partial x^{\mu}/\partial \bar{x}^{\nu} \right\rVert\), vanishes (\(J=0\)) at the throat \(\Sigma\). A well-known example of a branching transformation \(f\) is the Einstein–Rosen transformation \citep{einstein}, which transforms the one-sheeted Schwarzschild solution into a new two-sheeted state featuring wormhole geometry\footnote{Here, \(M\) is the mass, and \(r\) is the radial coordinate. Time and angular coordinates remain unchanged.}:
\begin{equation}
    r = u^2 + 2M.
\end{equation}In this state, the throat \(\Sigma\) is the sphere \(r=2M\), the Einstein–Rosen bridge.

Since the branching transformation \(f\) is not one-to-one, it generates a new spacetime configuration that differs from the original one in its topology. In this sense, topological dressing differs from a standard coordinate transformation: in the case of a standard coordinate transformation, we obtain merely a new description of the same underlying state. With topological dressing, however, we obtain a new state. The difference lies in the value of the Jacobian \(J\): in the case of a standard coordinate transformation, \(J \ne 0\)  throughout the entire space, whereas in the case of topological dressing, it vanishes (\(J=0\)) on a specific surface - the throat \(\Sigma\).

The working hypothesis of the topological dressing method is the following:

a vacuum solution may be extended into a two-sheeted spacetime by means of a branching transformation whose Jacobian vanishes on the throat surface \(\Sigma\)\footnote{The restriction to a vacuum is not mandatory; the case of an electrovacuum is also considered in \citep{dimaschko}.}.

In this sense, topological dressing suggests that the geometric content of Einstein covariance may extend beyond ordinary one-to-one coordinate transformations with \(J \ne 0\).

Since the metric determinant  \(g = \det\|g_{\mu \nu}\|\) transforms as \(\bar{g}=J^2 g\), the vanishing of the Jacobian (\(J=0\) ) implies that the metric of the new solution automatically becomes \textit{degenerate} at the throat \(\Sigma\):
\begin{equation}
    \bar{g} \big|_{\Sigma } =0.
\end{equation}Consequently, topological dressing transforms an already known one-sheeted solution of the vacuum Einstein equations into a two-sheeted solution describing a vacuum wormhole with a \textit{degenerate} metric.

Thus, the primary result of the topological dressing method is a two-sheeted degenerate metric describing a matter-free wormhole with an \textit{a priori} specified throat \(\Sigma\).

Let us enumerate the \textit{differences} between the thin-shell model and the topological dressing method:

i) In the thin-shell model, the throat \(\Sigma\) serves as a junction surface, whereas in the topological dressing method, it acts as a branching surface (see Fig. 1);

ii) In the thin-shell model, the throat \(\Sigma\) constitutes a distinct surface relative to the rest of spacetime; it is not described by the spacetime metric but is instead accounted for through Israel’s junction conditions. In the topological dressing method, the throat \(\Sigma\) is merely a part of the two-sheeted wormhole spacetime and is described by its metric;

iii) In the thin-shell model, the spacetime metric is \textit{non-degenerate}  (\(g \ne 0\)) throughout its entire domain of definition - which explicitly excludes the throat \(\Sigma\); in the topological dressing method, the resulting metric is \textit{degenerate}  (\(\bar{g}=0\)), with this degeneracy occurring precisely at the throat \(\Sigma\);

iv) In the thin-shell model, the throat \(\Sigma\) carries matter, which constitutes the primary result of this approach; the topological dressing method yields a complete two-sheeted metric describing a matter-free spacetime configuration.

It should be noted that, in the case of a vacuum, all these distinctions apply solely to the throat \(\Sigma\) - outside the throat, both approaches yield identical geometries. This raises the question: what is the physical significance of these differences? Does the topological dressing method truly yield a vacuum state throughout the entire two-sheeted spacetime - including the throat \(\Sigma\) - and is it indeed devoid of a material \(\delta\)-layer?

Since the topological dressing method automatically leads to a \textit{degenerate} metric, relying solely on the standard equations of General Relativity - derived from the Einstein-Hilbert action - is insufficient to answer this question. In the case of a \textit{degenerate} metric, the natural descriptive framework is the Einstein-Palatini-Cartan (EPC) tetrad formalism, governed by the action\footnote{Here, \(e^a\) is a tetrad vector, and \(R^{cd}\) is the curvature tensor as a function of the connections \( {\omega^a}_b \).}:
\begin{equation}
    S = -\frac{1}{32\pi} \int \varepsilon_{abcd}\, e^a \wedge e^b \wedge R^{cd}(\omega).
\end{equation}Just as Israel’s boundary conditions constitute an integral element of the thin-shell model - determining the matter residing at the throat of a \textit{non-degenerate} wormhole - the EPC formalism serves as a key element of the topological dressing method, demonstrating the absence of matter at the throat of a \textit{degenerate} wormhole.

\textit{In this work - without ruling out the possibility of proving this in a more general form - we simply demonstrate the absence of matter using three specific examples of a degenerate wormhole. In each case, we also examine a close non-degenerate geometry, showing how it reproduces the results of the thin-shell model and leads to the emergence of matter at the throat.}

\section{Example A: Rindler wormhole}

\textit{In this section, we consider a degenerate wormhole with planar throat. This configuration is based on a two-sheeted Rindler-type metric, which is obtained by topologically dressing a one-sheeted Rindler metric}

Consider the Rindler metric
\begin{equation}
    ds^{2} = e^{2 \alpha x} \left( d\tau^{2} - dx^{2} \right) - dy^{2} - dz^{2},
\end{equation}where \(-\infty < \tau, x, y, z < \infty\) and \(\alpha > 0\) is a constant parameter. In its standard interpretation, this metric describes a uniformly accelerated frame in flat spacetime \citep{rindler,MTW}. Our goal is to construct a two-sheeted configuration with a \textit{degenerate} throat starting from metric (3.1), and to determine whether such a configuration requires matter.

To obtain a two-sheeted wormhole metric from the one-sheeted metric (3.1), we perform the following branching coordinate transformation: 
\begin{equation}
    x = \sqrt{l^2 + a^2}.
\end{equation}As a result, the metric takes the form 
\begin{equation}
    ds^{2} = e^{2\alpha x}\left(d\tau^{2} - \frac{l^{2}}{x^{2}}\,dl^{2}\right) - dy^{2} - dz^{2},
\end{equation}where \(x\) is now expressed in terms of \(l\) according to (3.2). The new coordinate \(l \in  (- \infty, \infty)\) describes two sheets of the wormhole spacetime (see Fig. 2). Its throat is the surface \(l=0\), which corresponds to plane \(x=a\) in the one-sheeted spacetime. The determinant of the metric vanishes at \(l=0\), that is the wormhole is \textit{degenerate}. Further we refer to (3.3) as a Rindler wormhole with a planar throat \(\Sigma \equiv \{x=a \}\).

\begin{figure}[t!]
  \centering
    \includegraphics[scale=0.38]{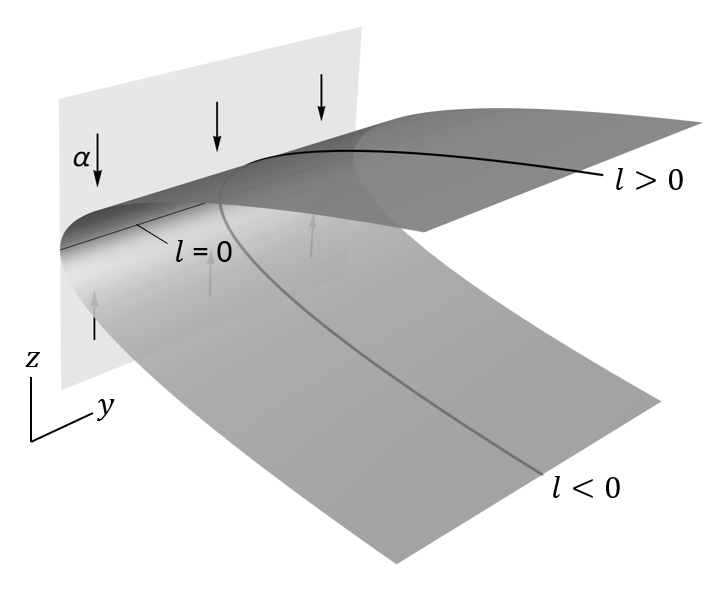} 
2  \caption{The vicinity of the throat of a \textit{degenerate} Rindler wormhole. The coordinate line of the two-sheeted coordinate \(l\) is shown. The coordinate value \(l = 0\) corresponds to the hypersurface of the \textit{degenerate} throat. The upper sheet of the wormhole is the region \(l > 0\), the lower sheet is the region \(l < 0\). In the tangent flat space with coordinates \((y,z)\), an observer located at the throat \(l = 0\) registers a gravitational field with the Rindler acceleration of gravity \(\alpha\), directed from both sides toward the throat.}
\end{figure}

Since the Rindler wormhole is \textit{degenerate}, it requires a tetrad description. Based on metric (3.3), we choose a natural coframe of the form 
\begin{equation}
    e^0 = e^{\alpha x} \, d\tau, \quad e^1 = e^{\alpha x} \, \frac{l}{x} \, dl, \quad e^2 = dy, \quad e^3 = dz.
\end{equation}
Then from zero torsion condition
\begin{equation}
    de^{i} + \omega^{i}_{\ j} \wedge e^{j} = 0
\end{equation}
the only non-zero connection is

\begin{equation}
    \omega^{01} = \alpha d\tau ,
\end{equation}and all curvature components vanish, 
\begin{equation}
    R^{ij} = d\omega^{ij} + \omega^{ik} \wedge \omega^{j}{}_{k} = 0. 
\end{equation}
Thereby, the coframe (3.4) is a solution to the vacuum Einstein equations in the first-order formulation. This signifies the absence of matter throughout the entire two-sheeted space. In particular, the surface mass density  at the throat is equal to zero:
\begin{equation}
    \sigma_m = 0.
\end{equation}Nevertheless, the absence of matter in this case does not imply the absence of gravitational effects.

In accordance with (3.6), the coordinate acceleration of a test particle initially at rest is equal to \( -\alpha\).\footnote{The exact equation of motion for a test particle is  \( d^{2} x / d\tau^{2} = -\alpha \left[ 1 - \left( dx / d\tau \right)^{2} \right]\)}  Here, it is important that the parameter \(\alpha\) in the two-sheeted Rindler wormhole metric (3.3) no longer has the same status as in the one-sheeted Rindler metric (3.1). In the ordinary Rindler spacetime it characterizes a non-inertial frame and can be removed by passing to inertial Minkowski coordinates. In the two-sheeted geometry (3.3), however, the coframe one-form \(e^1\) changes orientation when crossing the throat: on the upper sheet \(l>0\) it is directed toward increasing \(l\), whereas on the lower sheet \(l<0\) it is directed toward decreasing \(l\). Hence the same coordinate acceleration \(-\alpha\) corresponds to physical accelerations directed toward the surface \(l=0\) on both sheets. Static observers on both sides therefore register a gravitational acceleration of magnitude  \(|\alpha|\)  (or simply \(\alpha\) as \(\alpha >0\)) directed toward the throat. It is shown in Fig. 2.

Thus, the two-sheeted topology turns the removable acceleration of the one-sheeted Rindler frame into an invariant feature of the \textit{degenerate} configuration: a gravitational acceleration \(\alpha\) toward the throat, which persists even though the curvature and the matter content vanish. The cause of this gravitational effect is not the local geometry of spacetime (the Ricci curvature generated by matter), but rather its global topology (manifesting in the presence of a \textit{degenerate} throat).

Next, we compare the \textit{degenerate} Rindler metric with similar \textit{non-degenerate} metrics.

To this end, following the idea of Einstein and Rosen, we modify the metric (3.3) in such a way as to remove the degeneracy.The modified metric has the form
\begin{equation}
    ds^{2} = e^{2 \alpha x} \left( d\tau^{2} - \frac{l^{2} + \epsilon^{2}}{x^{2}} \, dl^{2} \right) - dy^{2} - dz^{2}.
\end{equation}
At \(\epsilon = 0\), the modified metric (3.9) becomes the \textit{degenerate} metric (3.3). At \(\epsilon \ne 0\), the metric (3.9) is \textit{non-degenerate} for any value of the global coordinate \(l\). However, it no longer satisfies  the vacuum Einstein equations and therefore requires matter. This leads to non-zero values of curvatures \(R^{ij}\), as well as of the action (2.4) and the resulting matter mass.

To calculate the action (2.4) in a space with the modified metric (3.9), we choose a \textit{different} orthonormal coframe
\begin{equation}
    e^0 = e^{\alpha x} \, d\tau, \quad e^1 = \frac{e^{\alpha x}}{x} A(l,\epsilon) \, dl, \quad e^2 = dy, \quad e^3 = dz.
\end{equation}Here, the function
\begin{equation}
    A(l,\epsilon) = \sqrt{l^{2} + \epsilon^{2}}
\end{equation}is differentiable in the case of a \textit{non-degenerate} metric (\(\epsilon \ne 0\)) that interests us. From (3.5) this yields a unique non-zero connection
\begin{equation}
    \omega^{01} = A'(l,\epsilon)\, d\tau
\end{equation}and  from (3.7) a unique non-zero curvature
\begin{equation}
    R^{01} = d\omega^{01} =A''(l,\epsilon)\,dl \wedge d\tau.
\end{equation}
Here, the prime means the derivative with respect to the coordinate \(l\) . Substituting this curvature into (2.4) yields
\begin{equation}
    S(\epsilon) = -\frac{\alpha}{8\pi} \int A''(l,\epsilon)\, d^{4}x.
\end{equation}The dependence on the parameter \(a\), which determines the \(x\)-coordinate of the throat, drops out of the action - as it should be due to the translational invariance of the system along the \(x\)-axis.

After separating the time integration in the action (3.14)\footnote{The sign convention differs from the standard particle action  \(S_{m} = -\int m \, d\tau\). In the present context, (3.15) effectively represents the \textit{gravitational} contribution rather than a conventional \textit{matter} action.}
\begin{equation}
    S(\epsilon) = \int m(\epsilon) \, d\tau
\end{equation}we can determine the mass \(m(\epsilon)\):
\begin{equation}
    m(\epsilon) = -\frac{\alpha}{8\pi} \int A''(l,\epsilon)\, dl\, dy\, dz ,
\end{equation}as well as the surface mass density \( \sigma_m (\epsilon)\)
\begin{equation}
    \sigma_m (\epsilon) = -\frac{\alpha}{8\pi} \int_{-\infty}^{\infty} A''(l,\epsilon)\, dl = -\frac{\alpha}{8\pi} \left[ A'(\infty,\epsilon) - A'(-\infty,\epsilon) \right].
\end{equation}It follows that in the case of a \textit{non-degenerate} state (\(\epsilon  \ne 0\)), when  \(A'(\pm\infty,\epsilon) = \pm 1\), we arrive at a non-zero value of the surface mass density
\begin{equation}
     \sigma_m (\epsilon) = -\frac{\alpha}{4\pi}, \quad (\epsilon \neq 0).
\end{equation} Combining (3.8) and (3.18), we obtain a complete expression that describes both the \textit{non-degenerate} and the \textit{degenerate} throat: 
\begin{equation}
    \sigma_m (\epsilon) =
    \begin{cases}
        -\dfrac{\alpha}{4\pi}, & \epsilon \neq 0 , \\
        \quad 0, & \epsilon = 0.
    \end{cases}
\end{equation}Thus, in the case of a \textit{non-degenerate} throat (\(\epsilon \ne 0\)), there is exotic matter on its surface with a negative surface mass density \( \sigma_m (\epsilon) = - \alpha /(4 \pi)\). In the case of a \textit{degenerate} throat (\(\epsilon = 0\)), there is no matter.

Referring to the integral representation (3.17) and introducing the volume density
\begin{equation}
     \rho_m (l,\epsilon) = -\frac{\alpha}{8\pi} A''(l,\epsilon) = -\frac{\alpha}{8\pi} \frac{\epsilon^{2}}{\left(l^{2} + \epsilon^{2}\right)^{3/2}} ,
\end{equation}both cases can be covered in terms of  volume density \(\rho_m (l,\epsilon)\):
\begin{equation}
    \rho_m (l,\epsilon) =
    \begin{cases}
        -\dfrac{\alpha}{8\pi} \epsilon^{2} \left(l^{2} + \epsilon^{2}\right)^{-3/2}, & \epsilon \neq 0,\\[6pt]
        \quad 0, & \epsilon = 0.
    \end{cases}
\end{equation} In the limit \(\epsilon \rightarrow 0\), the volume density \(\rho_m (l,\epsilon)\) reduces to the \(\delta\)-distribution
\begin{equation}
    \lim_{\epsilon \to 0} \rho_m (l,\epsilon) = -\frac{\alpha}{4\pi} \, \delta(l).
\end{equation} 
This \(\delta\)-distribution coincides with the result of the thin-shell model \citep{visser}. At the same time, \textit{at the very point} \(\epsilon = 0\), the function \(\rho_m (l,\epsilon)\) is identically zero:
\begin{equation}
    \rho_m (l,0) = 0.
\end{equation}

\begin{figure}[t!]
  \centering
    \includegraphics[scale=0.52]{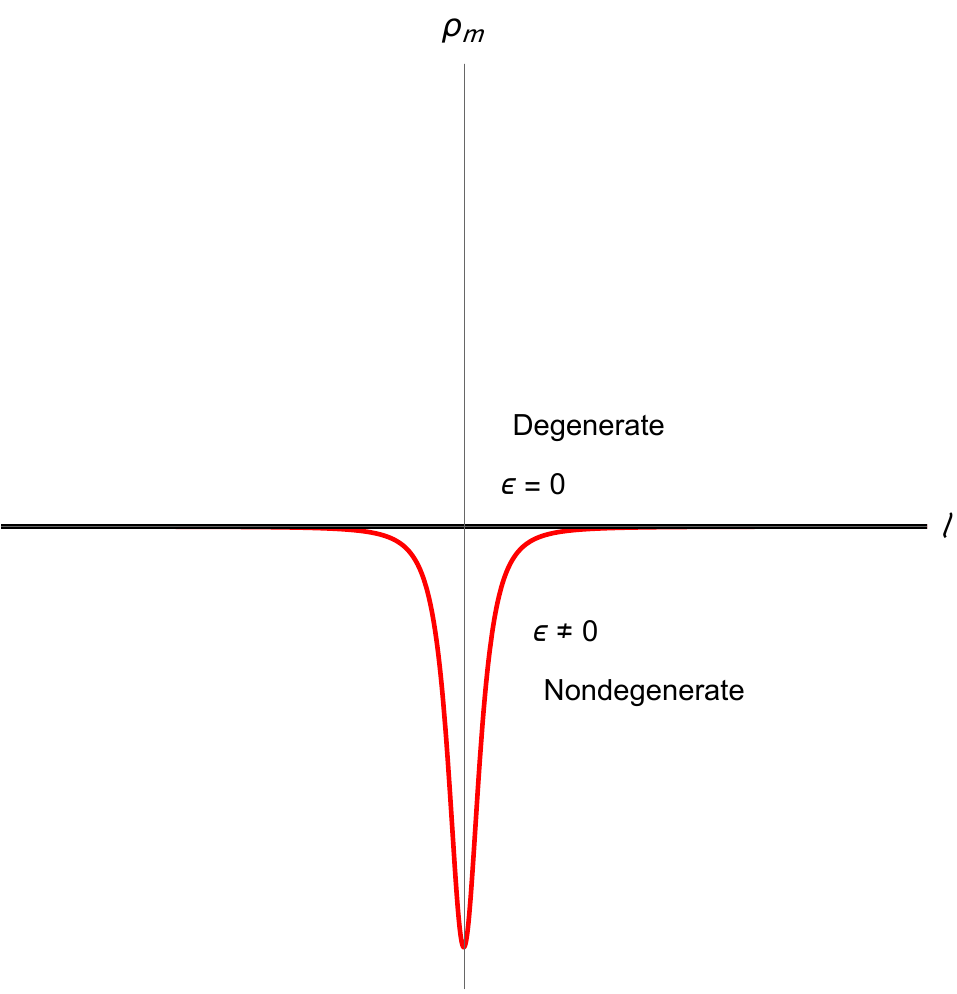} 
  \caption{Volumetric mass density of matter \(\rho_m (l,\epsilon)\) in the vicinity of a \textit{degenerate} throat \(\epsilon = 0\), black line) and a \textit{non-degenerate} throat (\(\epsilon \ne 0\), red line) in the Rindler metric. The limit \(\epsilon \rightarrow 0\) corresponds to the thin-shell model.}
\end{figure}

A significant point  is the discrepancy between the limiting density distribution \(\rho_m (l,\epsilon)\) as \(\epsilon \rightarrow 0\)  and the distribution at \(\epsilon = 0\):
\begin{equation}
    \lim_{\epsilon \to 0}\rho_m (l,\epsilon) \neq \rho_m (l,0).
\end{equation}In the first case (\textit{non-degenerate} throat), this is a \(\delta\)-function; in the second case (\textit{degenerate} throat), this is zero. Both cases are graphically represented in Fig. 3.

The reason for this discrepancy is \textit{not the formal vanishing }of the density \(\rho_m (l,\epsilon)\) according to (3.21) at \(\epsilon = 0\). This fact is \textit{not essential}, since the factor \(\epsilon^2\) is absorbed by the integration over \(dl\) in (3.16) of expression (3.20).

The reason for the discrepancy in (3.24) is that at \(\epsilon = 0\), the degeneracy factor  \( A(l,\epsilon) = \sqrt{l^2 + \epsilon^2}\) becomes a non-differentiable function of \(l\):  \( A(l,0)= \lvert l \rvert\). This makes it impossible to use the non-smooth coframe (3.10), in which \(e^1 \propto  \lvert l \rvert dl\), and requires returning to the smooth coframe (3.4) with \(e^1 \propto ldl\).

\textit{To summarize the findings of this section, we have shown that:}

\textit{- A Rindler wormhole with a degenerate planar throat is matter-free and generates no curvature;}

\textit{- nevertheless, this wormhole induces a gravitational acceleration - unrelated to the curvature - which is attributable to the presence of the degenerate throat;}

\textit{- a non-degenerate Rindler wormhole, infinitesimally close to the degenerate case, requires the presence of exotic matter at the throat; this precisely reproduces the \(\delta\)-layer of the thin-shell model in this geometry.}

\section{Example B: Klinkhamer wormhole}

\textit{\textit{In this section, we consider a degenerate wormhole with a spherical throat that generates no gravitational effects. This configuration is based on the two-sheeted Klinkhamer metric, obtained through the topological dressing of Minkowski spacetime.}
}

We consider the topological dressing of flat Minkowski spacetime, which has the following metric in spherical coordinates \footnote{The usual notations for spherical coordinates \((r, \theta, \varphi)\) are adopted, \(d\Omega^{2} = d\theta^{2} + \sin^{2}\theta \, d\varphi^{2}\) is the metric element of the solid angle.}
\begin{equation}
    ds^{2} = dt^{2} - dr^{2} - r^{2} d\Omega^{2}.
\end{equation}As the branching transformation, we select the following transformation of the radial coordinate \(r\):
\begin{equation}
    r = \sqrt{l^2 + a^2}, \quad a > 0.
\end{equation}
Here, we introduce a two-sheeted coordinate  \( l \in (-\infty, \infty)\) and a positive parameter \(a\), which represents the radius of the spherical throat. Under this transformation, the Minkowski metric takes the form of the Klinkhamer metric,
\begin{equation}
    ds^2 = dt^2 - \frac{l^2}{r^2} \, dl^2 - r^2 \, d\Omega^2.
\end{equation}The determinant of this metric is
\begin{equation}
    g = -l^2 (l^2 + a^2) \sin^2 \theta
\end{equation}which vanishes on the surface \(l=0\) or, equivalently, on the sphere \(r=a\). This determines the location of the throat hypersurface \(\Sigma \equiv \{ r = a \}\).  Further we refer to (4.3) as \textit{Klinkhamer wormhole.} As its metric (4.3) is \textit{degenerate}, this necessitates a tetradic description.

Based on metric (4.3), we choose a coframe of the form  
\begin{equation}
    e^0 = dt, \quad e^1 = \frac{l}{r} \, dl, \quad e^2 = r \, d\theta, \quad e^3 = r \sin\theta \, d\varphi .
\end{equation}Then, three non-zero connections are
\begin{equation}
    \omega^{12} = -d\theta, \qquad \omega^{13} = -\sin\theta\, d\varphi, \qquad \omega^{23} = -\cos\theta\, d\varphi ,
\end{equation}which, after substituting into (3.7), vanish all curvature components:
\begin{equation}
    R^{ij} = 0.
\end{equation}
Thereby, the coframe (4.5) is a solution to the vacuum Einstein equations in the first-order formulation. This signifies the absence of matter throughout the entire two-sheeted space and, accordingly, the mass of matter is zero. However, in contrast to the Rindler wormhole, all connection components involving the time direction also vanish:
\begin{equation}
    \omega^{01} = \omega^{02} = \omega^{03} = 0.
\end{equation}Consequently, static observers experience no gravitational acceleration like that in the Rindler case: the Klinkhamer wormhole possesses neither tidal effects nor local gravitational manifestations. Therefore, this \textit{degenerate} structure is purely topological and generates no gravitational effects.

Next, we compare the degenerate Klinkhamer metric with similar \textit{non-degenerate} metrics.
To eliminate the singularity at the throat, we modify the degenerate metric (4.3) to remove the degeneracy. The modified metric has the form
\begin{equation}
    ds^{2} = dt^{2} - \frac{l^{2} + \epsilon^{2}}{r^{2}}\,dl^{2} - r^{2}\,d\Omega^{2}
\end{equation}and the corresponding modified coframe
\begin{equation}
    e^0 = dt, \quad e^1 = \frac{\sqrt{l^2 + \epsilon^2}}{r} \, dl, \quad e^2 = r \, d\theta, \quad e^3 = r \sin\theta \, d\varphi .
\end{equation}At \( \epsilon =0\), the modified metric (4.9) becomes the \textit{degenerate} metric (4.3). At \( \epsilon \ne 0\), it remains \textit{non-degenerate}, as the determinant of metric (4.9)
\begin{equation}
    g = - (l^2 + \epsilon^2)(l^2 + a^2)\sin^2{\theta} 
\end{equation}does not vanish anywhere except at coordinate singularities \(\theta = 0, \pi\). In this case, metric (4.9) does not satisfy Einstein's equations in a vacuum and requires matter. This means that the curvature \(R^{ij}\) and the action (2.4) are non-zero. According to the general expression (3.15), this corresponds to a non-zero matter mass \(m(\epsilon)\).

The matter mass is determined by the curvature scalar \(R(l,\epsilon)\) corresponding to a given value of the parameter \(\epsilon\). Its expression is given  in Appendix A:
\begin{equation}
   R(l,\epsilon) = \frac{2 \epsilon^{2}}{l^{2} + \epsilon^{2}} \left( \frac{1}{l^{2} + a^{2}} - \frac{2}{l^{2} + \epsilon^{2}} \right).
\end{equation}
Based on the general expression (3.15) for single-particle action, the mass of matter \(m(\epsilon)\) is expressed by the integral over the entire 3D volume of a two-sheeted space 
\begin{equation}
    m(\epsilon) = \frac{1}{16 \pi} \int R(l,\epsilon) \, dV.
\end{equation}
For a \textit{non-degenerate} state with metric (4.9), the volume measure is\begin{equation}
    dV = \sqrt{(l^2 + \epsilon^2)(l^2 + a^2)} \, d\Omega \, dl,
\end{equation}which reduces expression (4.13) to an integral over the two-sheeted coordinate \(l\),
\begin{equation}
    m(\epsilon) = \frac{\epsilon^2}{2} \int_{-\infty}^{\infty} \left( \frac{1}{l^2 + a^2} - \frac{2}{l^2 + \epsilon^2} \right) \sqrt{\frac{l^2 + a^2}{l^2 + \epsilon^2}} \, dl.
\end{equation}
The asymptotic behavior of the mass  \(m(\epsilon)\) in the limit \(\epsilon \rightarrow 0\) is (see Appendix A)
\begin{equation}
    \lim_{\epsilon \to 0} m(\epsilon) = -2a, 
\end{equation}which corresponds to the limiting value of the surface mass density
\begin{equation}
    \lim_{\epsilon \to 0} \sigma_m (\epsilon) = -\frac{1}{2\pi a}.
\end{equation}
Since both expressions are negative, the matter stabilizing the  \textit{non-degenerate} throat must be exotic. Relations (4.16) and (4.17) give, respectively, the integral mass \(m(\epsilon)\) and the surface mass density \(\sigma_m (\epsilon)\) of the exotic matter necessary to stabilize a wormhole with \textit{non-degenerate} metric (4.9), taken in the limit of a sufficiently thin throat (in this case, this means that the parameter \(\epsilon/a \ll 1\) is small).

It is appropriate  to compare the limit expression (4.17) with the result of the thin-shell model for a wormhole, obtained by matching two flat Minkowski spacetimes over a sphere of radius \(a\) \citep{visser}. In this case, the thin-shell model yields the stress-energy tensor
\begin{equation}
    T_{\nu}^{\mu} = \operatorname{diag}(\sigma, 0, p, p)\,\delta(l),
\end{equation}
where
\begin{equation}
    \sigma = -\frac{1}{2\pi a}, \qquad p = -\frac{1}{4\pi a}.
\end{equation}
 We use the expression for the action following from Einstein's equations\begin{equation}
    S = \frac{1}{2} \int T_{\mu}^{\mu} \sqrt{-g} \, d^{4}x .
\end{equation}
Then a comparison with (3.15) yields the surface mass density \(\sigma_{m}\) in the variational approach. For the thin-shell model, its magnitude can be expressed in terms of the components of the tensor \( T_{\nu}^{\mu}\) as\begin{equation}
    \sigma_m = \frac{1}{2}(\sigma + 2p).
\end{equation}After substituting from (4.19), this yields
\begin{equation}
    \sigma_m = -\frac{1}{2\pi a},
\end{equation}
which coincides with the limit expression for \(\sigma_m(\epsilon)\) as  \(\epsilon \rightarrow 0\)  according to (4.17). The coincidence of these expressions shows that the thin-shell model naturally emerges as the limit of a \textit{non-degenerate} regularization of a \textit{degenerate} geometry.

As in the previous example, here the limiting value of the surface mass density \(\sigma_m(\epsilon)\) as  \(\epsilon \rightarrow 0\)  (non-degenerate geometry) also differs from the zero value of this same quantity at \(\epsilon = 0\) (degenerate geometry).

\textit{To summarize the findings of this section, we have shown that:}

\textit{- A degenerate Klinkhamer wormhole is matter-free and generates no gravitational effects;}

\textit{- A non-degenerate Klinkhamer wormhole, infinitesimally close to the degenerate case, requires the presence of exotic matter at the throat; this precisely reproduces the \(\delta\)-layer of the thin-shell model in this geometry.}

\section{Example C: Schwarzschild-Klinkhamer wormhole }

\textit{\textit{In this section, we consider a degenerate wormhole with a spherical throat that generates gravitational effects. This configuration is based on the two-sheeted Schwarzschild-Klinkhamer metric, obtained through the topological dressing of Schwarzschild spacetime} .}

We consider the topological dressing of Schwarzschild spacetime, which has following metric in spherical coordinates
\begin{equation}
    ds^2 = \left(1 - \frac{2M}{r}\right) dt^2 - \left(1 - \frac{2M}{r}\right)^{-1} dr^2 - r^2 d\Omega^2
\end{equation}As the branching transformation, we select the following transformation of the radial coordinate   \(r\):
\begin{equation}
    r = \sqrt{l^2 + a^2}, \quad a > 2M. 
\end{equation}We introduce a two-sheeted coordinate \( l \in (-\infty, \infty)\) and a positive parameter \(a\), which represents the radius of the spherical throat. After this transformation, the Schwarzschild metric becomes the Schwarzschild-Klinkhamer metric 
\begin{equation}
    ds^{2} = \left(1 - \frac{2M}{r}\right) dt^{2}
    - \left(1 - \frac{2M}{r}\right)^{-1} \frac{l^{2}}{r^{2}} \, dl^{2}
    - r^{2} d\Omega^{2}.
\end{equation}
It has a determinant
\begin{equation}
    g = -l^2 (l^2 + a^2)\sin^2 \theta
\end{equation}which vanishes on the surface \(l=0\). This determines the position of the throat surface, \(\Sigma \equiv \{ r = a \}\). Further we refer to (5.3) as \textit{Schwarzschild-Klinkhamer wormhole}. As metric (5.3) is \textit{degenerate}, this necessitates a tetradic description.

Based on metric (5.3), we choose a coframe of the form 
\begin{equation}
    e^{0} = f \, dt, \quad e^{1} = \frac{l}{rf} \, dl, \quad e^{2} = r \, d\theta, \quad e^{3} = r \sin\theta \, d\varphi,
\end{equation}or, equivalently,
\begin{equation}
    e^0 = f \, dt, \quad e^1 = \frac{dr}{f}, \quad e^2 = r \, d\theta, \quad e^3 = r \sin\theta \, d\varphi.
\end{equation}Here we introduce the notation
\begin{equation}
    f = \sqrt{1 - \frac{2M}{r}}.
\end{equation}
The tetrad of the \textit{degenerate} state in the form (5.6) reproduces the Schwarzschild metric (5.1) for all values of \(r \ge a>2M\). Since the Schwarzschild metric (5.1) is a solution of the Einstein equations in a vacuum for all \(r>2M\), the tetrad (5.6) also gives a vanishing Ricci tensor on the entire two-sheeted space of the wormhole, \( -\infty < l < \infty\). \footnote{The tetrad calculation of the curvature tensor proceeds exactly as in the ordinary Schwarzschild geometry and therefore will not be repeated here (see, e.g., \citep{MTW,carroll}).}

Consequently, the \textit{degenerate} metric (5.3), which describes the Schwarzschild-Klinkhamer wormhole, does not require matter anywhere, including at the throat. In particular, this is manifested in the fact that the mass of matter, defined through action (2.4 ) according to (3.15), also vanishes.

Despite the absence of matter in the entire two-sheeted space (including the throat), the \textit{degenerate} Schwarzschild-Klinkhamer wormhole is the source of the gravitational field. This can be verified by the non-zero value of the Weyl invariant. Due to the local coincidence of tetrads (5.5) and (5.6) for the Schwarzschild and Schwarzschild-Klinkhamer metrics, in the case of a degenerate Schwarzschild-Klinkhamer wormhole, the Weyl invariant has the same form as for the Schwarzschild metric: 
\begin{equation}
    C_{\alpha\beta\gamma\delta} C^{\alpha\beta\gamma\delta} = \frac{48 M^{2}}{r^{6}},
\end{equation}where, according to relation (5.2), the denominator contains the non-zero expression:  \(r^{6} = \left(l^{2} + a^{2}\right)^{3}\). Thus, the gravitational field extends throughout the entire two-sheeted spacetime and has no singularities at the throat  \(l=0\). 

It should also be noted that near the throat \(l=0\), the geometry of the Schwarzschild-Klinkhamer metric becomes close to the geometry of the Rindler metric; see Fig. 4. This makes the distribution of the gravitational field in this region clear - for observers at the throat, it is uniform. Near the throat \(r=a\), the Schwarzschild geometry locally approaches a Rindler-type configuration with effective acceleration parameter \(\alpha=M/a^2\).

\begin{figure}[b!]
  \centering
    \includegraphics[scale=0.55]{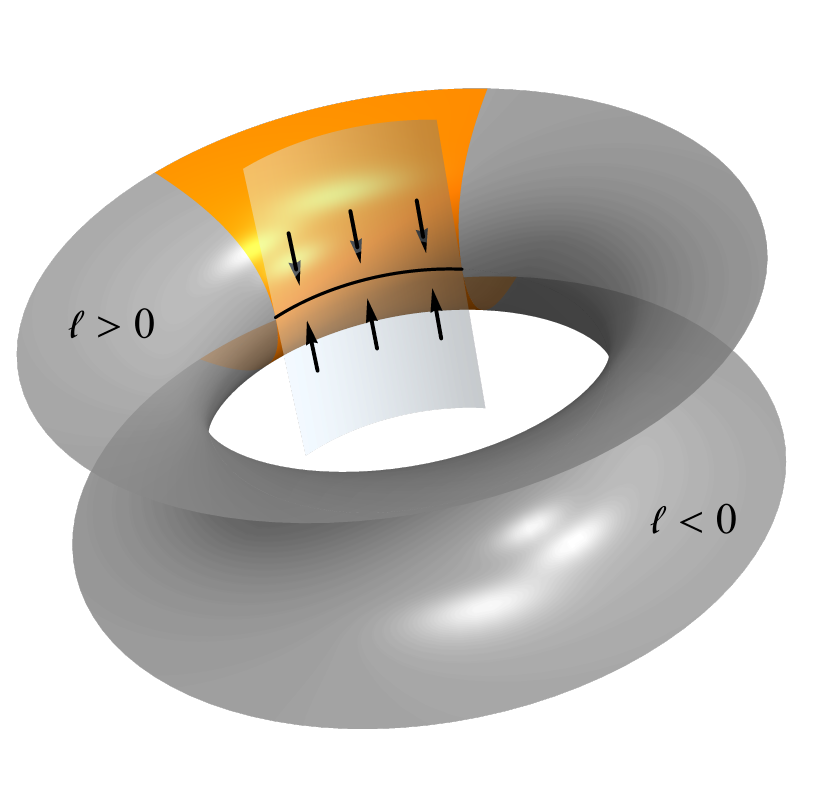} 
  \caption{Degenerate Schwarzschild-Klinkhamer wormhole with upper sheet \(l>0\) and lower sheet \(l<0\). Its locally Rindler-like fragment is highlighted in orange, cf. Fig. 1. In the tangent space, an observer located at the throat registers a gravitational field with Rindler acceleration of gravity \(\alpha =M/a^2\), directed from both sides toward the throat.}
\end{figure}

Using the same method as in the previous sections, we can eliminate the degeneracy at the throat. To do this, it suffices to modify the \textit{degenerate} metric (5.3) as follows:

\begin{equation}
    ds^2 = \left(1 - \frac{2M}{r}\right) dt^2 - \left(1 - \frac{2M}{r}\right)^{-1} \frac{l^2 + \epsilon^2}{r^2} \, dl^2 - r^2 d\Omega^2.
\end{equation} For \(\epsilon = 0\), the modified metric (5.9) becomes the \textit{degenerate} metric (5.3). For \(\epsilon \ne 0\), it remains \textit{non-degenerate}, since the determinant of metric (5.9)
\begin{equation}
    g = - (l^2 + \epsilon^2)(l^2 + a^2)\sin^2 \theta
    \label{eq:placeholder_label}
\end{equation}does not vanish anywhere except at coordinate singularities \(\theta = 0, \pi\). However, in this case, the metric (5.9) does not satisfy Einstein's equations in vacuum and requires matter.

For the \textit{non-degenerate} state (5.9) with \(\epsilon \ne 0\), we choose another orthonormal coframe
\begin{equation}
    e^0 = f\,dt, \quad e^1 = \frac{\sqrt{l^2 + \epsilon^2}}{f r}\,dl, \quad e^2 = r\,d\theta, \quad e^3 = r \sin\theta\,d\varphi.
\end{equation}In this case, the curvature and the action are nonzero, which, according to the general expression (3.15), corresponds to a nonzero mass of matter \(m(\epsilon)\). 

This mass is determined by the curvature scalar \(R(l,\epsilon)\) corresponding to a given value of the parameter \(\epsilon\). Its value is found in Appendix A:
\begin{equation}
    R(l,\epsilon) = \frac{2\epsilon^{2}}{l^{2} + \epsilon^{2}} \left[ \frac{1}{l^{2} + a^{2}} - \frac{2}{l^{2} + \epsilon^{2}} + \frac{3M}{\sqrt{l^{2} + a^{2}} \left(l^{2} + \epsilon^{2}\right)} \right].
\end{equation}Based on the general expression (3.15) for single-particle action, the mass of matter  \(m(\epsilon)\) is expressed by the integral over the entire three-dimensional volume of the two-sheeted space
\begin{equation}
    m(\epsilon) = \frac{1}{16\pi} \int R(l,\epsilon) \, dV.
\end{equation}For a \textit{non-degenerate} state with metric (5.9), the volume measure is
\begin{equation}
    dV = \frac{\sqrt{\left(l^2 + \epsilon^2\right)\left(l^2 + a^2\right)}}{\sqrt{1 - \frac{2M}{r}}} \, d\Omega \, dl, 
\end{equation}which reduces expression (5.13) to an integral over the two-sheeted coordinate \(l\)

\begin{equation}
\begin{aligned}
m(\epsilon)
&= \frac{\epsilon^{2}}{2}
   \int_{-\infty}^{\infty}
   \Bigg[
     \frac{1}{l^{2} + a^{2}}
     - \frac{2}{l^{2} + \epsilon^{2}}
     + \frac{3M}{\sqrt{l^{2} + a^{2}} (l^{2} + \epsilon^{2})}
   \Bigg] \\
&\quad \times
   \sqrt{\frac{l^{2} + a^{2}}{l^{2} + \epsilon^{2}}}
   \frac{dl}{\sqrt{1 - \frac{2M}{r}}}.
\end{aligned}
\end{equation}The asymptotics of the integral mass  \(m(\epsilon)\) in the limit \(\epsilon \rightarrow 0\) has the form (see Appendix B)
\begin{equation}
    \lim_{\epsilon \to 0} m(\epsilon) = -\frac{2a - 3M}{\sqrt{1 - \frac{2M}{a}}},
\end{equation}which corresponds to the limiting value of the surface mass density
\begin{equation}
    \lim_{\epsilon \to 0} \sigma(\epsilon) = -\frac{1}{2\pi a} \frac{1 - \frac{3}{2}\frac{M}{a}}{\sqrt{1 - \frac{2M}{a}}}.
\end{equation}Taking into account expression (4.21) for the surface mass density through the stress-energy tensor, this value exactly reproduces the result of the thin-shell model \citep{visser}.

As in both previous  examples, here the limiting value of the surface mass density \(\sigma_m(\epsilon)\) as  \(\epsilon \rightarrow 0\)  (\textit{non-degenerate} geometry) also differs from the zero value of this same quantity at \(\epsilon = 0\) (\textit{degenerate} geometry).

\textit{To summarize the findings of this section, we have shown that:}

\textit{- A degenerate Schwarzschild-Klinkhamer wormhole is matter-free;}

\textit{- nevertheless, this wormhole generates a Schwarzschild gravitational field throughout the two-sheeted spacetime;}

\textit{- a non-degenerate Schwarzschild-Klinkhamer wormhole, infinitesimally close to the degenerate case, requires the presence of exotic matter at the throat; this precisely reproduces the result of the thin-shell model in this geometry.}

\section{Conclusions}

In this work, we have examined three examples of degenerate spacetime geometries with wormhole topology. All three configurations were constructed within a unified framework based on the topological dressing method \citep{dimaschko}, which applies branching coordinate transformations to known vacuum solutions of Einstein’s equations. In all cases, the Einstein–Palatini–Cartan formalism—which remains well defined in the presence of degeneracy—confirms the absence of matter throughout the entire two-sheeted spacetime, including the throat. In this regard, these \textit{degenerate} configurations differ fundamentally from their \textit{non-degenerate} counterparts in the thin-shell model, which require exotic matter \citep{visser}.

Despite the absence of matter, the three configurations exhibit distinct physical manifestations:

A) the \textit{degenerate} Rindler wormhole with a planar throat produces a gravitational acceleration effect of purely topological origin, while spacetime curvature remains absent; 

B) the \textit{degenerate} Klinkhamer wormhole, characterized by a vanishing mass parameter \(M\), possesses nontrivial topology but generates neither curvature nor gravitational effects; 

C) the \textit{degenerate} Schwarzschild–Klinkhamer wormhole, characterized by nonzero \(M\), generates nontrivial curvature together with a nonvanishing Weyl tensor associated with tidal gravitational effects. 

In all three examples, matter is absent only in the exactly \textit{degenerate} sector. Once degeneracy is lifted and the geometry becomes \textit{non-degenerate}, exotic matter reappears, reproducing in the limiting case the standard thin-shell description.

These results suggest that, within an extended configuration space, nontrivial spacetime geometry need not always be associated with conventional matter sources. In this sense, \textit{degenerate} geometries may be viewed as an independent sector of general relativity admitting matter-free configurations.

\appendix
\section{Action for the regularized Schwarzschild-Klinkhamer metric}
Let us write the regularized Schwarzschild-Klinkhamer metric (5.9) in a short form\begin{equation}
    ds^{2} = f^{2} dt^{2} - B^{2} dl^{2} - r^{2} d\Omega^{2},
\end{equation}where the notations are used
\begin{equation}
\begin{aligned}
r &= \sqrt{l^{2} + a^{2}}, &
\beta &= \sqrt{l^{2} + \epsilon^{2}}, \\
f &= \sqrt{1 - \frac{2M}{r}}, &
B &= \frac{\beta}{r f}.
\end{aligned}
\end{equation}Let us take the coframe
\begin{equation}
    e^0 = f \, dt, \quad e^1 = B \, dl, \quad e^2 = r \, d\theta, \quad e^3 = r \sin\theta \, d\varphi.
\end{equation}For this form, it is convenient to introduce functions
\begin{equation}
    u = \frac{1}{B} \frac{d (\ln f)}{d l} \,, \qquad v = \frac{1}{B} \frac{d (\ln r)}{d l} \,.
\end{equation}From the zero-torsion equation
\begin{equation}
    d e^{a} + \omega^{a}{}_{b} \wedge e^{b} = 0
\end{equation}follow components of connection:
\begin{equation}
    \omega^{0}{}_{1} = u e^{0}, \quad \omega^{2}{}_{1} = v e^{2}, \quad \omega^{3}{}_{1} = v e^{3}, \quad \omega^{3}{}_{2} = \frac{\cot \theta}{r} e^{3}
\end{equation}From the definition of the curvature
\begin{equation}
    R^{a}{}_{b} = d\omega^{a}{}_{b} + \omega^{a}{}_{c} \wedge \omega^{c}{}_{b}
\end{equation}we obtain its components
\begin{equation}
    \begin{aligned}
        R^{01} &= -\left(\frac{u'}{B} + u^{2}\right) e^{0} \wedge e^{1}, 
        &\;
        R^{02} &= -u v'' \, e^{0} \wedge e^{2}, \\
        R^{03} &= -u v'' \, e^{0} \wedge e^{3}, 
        &\;
        R^{12} &= -\left(\frac{v'}{B} + v^{2}\right) e^{1} \wedge e^{2}, \\
        R^{13} &= -\left(\frac{v'}{B} + v^{2}\right) e^{1} \wedge e^{3}, 
        &\;
        R^{23} &= \left(\frac{1}{r^{2}} - v^{2}\right) e^{2} \wedge e^{3},
    \end{aligned}
\end{equation}giving the curvature scalar
\begin{equation}
    R = 2\left(\frac{u'}{B} + u^2 + 2uv + \frac{2v'}{B} + 3v^2 - \frac{1}{r^2}\right).
\end{equation}Here, the prime denotes \(d/dl\). Taking into account \(r'=r/l\) and the definitions (\(A.2,A.4\)) we can express \(u\) and \(v\):
\begin{equation}
    u = \frac{Ml}{r^{2} \beta f}, \qquad v = \frac{fl}{r \beta}.
\end{equation}After substituting \(u\), \(v\) and \(B\) according to (\(A.10\)) and (\(A.2\)) into (\(A.9\)), the scalar \(R\) takes the form
\begin{equation}
\begin{aligned}
R &= \frac{2\epsilon^{2} + \beta^{2} - 2r^{2} + 3Mr}{\beta^{4} r^{2}} \\
  &= \frac{2\epsilon^{2}}{l^{2} + \epsilon^{2}}
     \Bigg[
       \frac{1}{l^{2} + a^{2}}
       - \frac{2}{l^{2} + \epsilon^{2}}
       + \frac{3M}{\sqrt{l^{2} + a^{2}}\,(l^{2} + \epsilon^{2})}
     \Bigg],
\end{aligned}
\end{equation}which gives the formula (5.12) of the main text.

\section{Thin-shell limit for the regularized Schwarzschild-Klinkhamer metric}
Consider expression (4.15) for the mass of matter in the \textit{non-degenerate} state of the Schwarzschild-Klinkhamer wormhole:
\begin{equation}
\begin{aligned}
m(\epsilon)
&= \frac{\epsilon^{2}}{2}
   \int_{-\infty}^{\infty}
   \Bigg[
     \frac{1}{l^{2} + a^{2}}
     - \frac{2}{l^{2} + \epsilon^{2}}
     + \frac{3M}{\sqrt{l^{2} + a^{2}} (l^{2} + \epsilon^{2})}
   \Bigg] \\
&\quad \times
   \sqrt{\frac{l^{2} + a^{2}}{l^{2} + \epsilon^{2}}}
   \frac{dl}{\sqrt{1 - \frac{2M}{r}}}.
\end{aligned}
\end{equation}It follows from the structure of the integral that in the limit \(\epsilon \rightarrow 0\):

\noindent 1) The first term in the square brackets gives a zero contribution to the integral;

\noindent 2) In the rest of the expression, except for the form \((l^2+\epsilon^2)\), we can set \(l=0\), which corresponds to the throat surface.

After this, the integral (B.1) takes the form of
\begin{equation}
   m(\epsilon)
    = -\frac{2a - 3M}{\sqrt{1 - \dfrac{2M}{a}}}
    \cdot \frac{\epsilon^{2}}{2}
    \int_{-\infty}^{\infty}
        \frac{\epsilon^{2} \, dl}{\left(l^{2} + \epsilon^{2}\right)^{3/2}},
\end{equation}or, after taking the integral,
\begin{equation}
    m(\epsilon) = -\frac{2a - 3M}{\sqrt{1 - \frac{2M}{a}}}.
\end{equation}The last expression is formula (5.16) of the main text.

\bibliographystyle{elsarticle-harv} 
\bibliography{main}

\begin{thebibliography}{20}
\expandafter\ifx\csname natexlab\endcsname\relax\def\natexlab#1{#1}\fi
\providecommand{\url}[1]{\texttt{#1}}
\providecommand{\href}[2]{#2}
\providecommand{\path}[1]{#1}
\providecommand{\DOIprefix}{doi:}
\providecommand{\ArXivprefix}{arXiv:}
\providecommand{\URLprefix}{URL: }
\providecommand{\Pubmedprefix}{pmid:}
\providecommand{\doi}[1]{\href{http://dx.doi.org/#1}{\path{#1}}}
\providecommand{\Pubmed}[1]{\href{pmid:#1}{\path{#1}}}
\providecommand{\bibinfo}[2]{#2}
\ifx\xfnm\relax \def\xfnm[#1]{\unskip,\space#1}\fi
%Type = Article
\bibitem[{Baines et~al.(2023)Baines, Gaur and Visser}]{baines}
\bibinfo{author}{Baines, J.}, \bibinfo{author}{Gaur, R.}, \bibinfo{author}{Visser, M.}, \bibinfo{year}{2023}.
\newblock \bibinfo{title}{Defect wormholes are defective}.
\newblock \bibinfo{journal}{Universe} \bibinfo{volume}{9}, \bibinfo{pages}{452}.
%Type = Book
\bibitem[{Carroll(2004)}]{carroll}
\bibinfo{author}{Carroll, S.M.}, \bibinfo{year}{2004}.
\newblock \bibinfo{title}{Spacetime and Geometry: An Introduction to General Relativity}.
\newblock \bibinfo{publisher}{Addison Wesley}, \bibinfo{address}{New York}.
%Type = Article
\bibitem[{Dimaschko(2024)}]{dimaschko}
\bibinfo{author}{Dimaschko, J.}, \bibinfo{year}{2024}.
\newblock \bibinfo{title}{Topological dressing method for the einstein-maxwell equations}.
\newblock \bibinfo{journal}{Gen. Relativ. Gravit.} \bibinfo{volume}{56}, \bibinfo{pages}{103}.
%Type = Article
\bibitem[{Dimaschko(2025)}]{dimaschko2}
\bibinfo{author}{Dimaschko, J.}, \bibinfo{year}{2025}.
\newblock \bibinfo{title}{Matter-free gravitational collapse and the equivalence principle}.
\newblock \bibinfo{journal}{Int. J. Geom. Methods Mod. Phys. (accepted), arxiv.org/abs/2512.16933} .
%Type = Article
\bibitem[{Einstein and Rosen(1935)}]{einstein}
\bibinfo{author}{Einstein, A.}, \bibinfo{author}{Rosen, N.}, \bibinfo{year}{1935}.
\newblock \bibinfo{title}{The particle problem in the general theory of relativity}.
\newblock \bibinfo{journal}{Phys. Rev.} \bibinfo{volume}{48}, \bibinfo{pages}{73--77}.
%Type = Article
\bibitem[{Feng(2023)}]{feng}
\bibinfo{author}{Feng, J.C.}, \bibinfo{year}{2023}.
\newblock \bibinfo{title}{Smooth metrics can hide thin shells}.
\newblock \bibinfo{journal}{Class. Quantum Grav.} \bibinfo{volume}{40}, \bibinfo{pages}{197002}.
%Type = Article
\bibitem[{Geroch(1967)}]{geroh}
\bibinfo{author}{Geroch, R.P.}, \bibinfo{year}{1967}.
\newblock \bibinfo{title}{Topology in general relativity}.
\newblock \bibinfo{journal}{J. Math. Phys.} \bibinfo{volume}{8}, \bibinfo{pages}{782--786}.
%Type = Article
\bibitem[{Hochberg and Visser(1998)}]{hochberg}
\bibinfo{author}{Hochberg, D.}, \bibinfo{author}{Visser, M.}, \bibinfo{year}{1998}.
\newblock \bibinfo{title}{Dynamic wormholes, antitrapped surfaces, and energy conditions}.
\newblock \bibinfo{journal}{Phys. Rev.} \bibinfo{volume}{D58}, \bibinfo{pages}{044021}.
%Type = Article
\bibitem[{Horowitz(1991)}]{horowitz}
\bibinfo{author}{Horowitz, G.T.}, \bibinfo{year}{1991}.
\newblock \bibinfo{title}{Topology change in classical and quantum gravity}.
\newblock \bibinfo{journal}{Class. and Quant. Gravit.} \bibinfo{volume}{8}, \bibinfo{pages}{587–602}.
%Type = Article
\bibitem[{Israel(1966)}]{israel}
\bibinfo{author}{Israel, W.}, \bibinfo{year}{1966}.
\newblock \bibinfo{title}{Singular hypersurfaces and thin shells m general relativity}.
\newblock \bibinfo{journal}{Nuovo Cimento} \bibinfo{volume}{44}, \bibinfo{pages}{1--14}.
%Type = Article
\bibitem[{Katanaev(2006)}]{katanaev}
\bibinfo{author}{Katanaev, M.O.}, \bibinfo{year}{2006}.
\newblock \bibinfo{title}{Polynomial hamiltonian form of general relativity}.
\newblock \bibinfo{journal}{Theoret. and Math. Phys.} \bibinfo{volume}{148}, \bibinfo{pages}{1264--1294}.
%Type = Article
\bibitem[{Klinkhamer(2023a)}]{klinkhamer1}
\bibinfo{author}{Klinkhamer, F.R.}, \bibinfo{year}{2023}a.
\newblock \bibinfo{title}{Defect wormhole: A traversable wormhole without exotic matter}.
\newblock \bibinfo{journal}{Acta Phys. Pol.} \bibinfo{volume}{B54}, \bibinfo{pages}{5--A3}.
%Type = Article
\bibitem[{Klinkhamer(2023b)}]{klinkhamer2}
\bibinfo{author}{Klinkhamer, F.R.}, \bibinfo{year}{2023}b.
\newblock \bibinfo{title}{Vacuum-defect wormholes and a mirror world}.
\newblock \bibinfo{journal}{Acta Phys. Pol.} \bibinfo{volume}{B54}, \bibinfo{pages}{7--22}.
%Type = Article
\bibitem[{Klinkhamer(2025)}]{klinkhamer3}
\bibinfo{author}{Klinkhamer, F.R.}, \bibinfo{year}{2025}.
\newblock \bibinfo{title}{Big bang as spacetime defect}.
\newblock \bibinfo{journal}{Mod. Phys. Lett. A} \bibinfo{volume}{40}, \bibinfo{pages}{2530010}.
%Type = Book
\bibitem[{Lichnerowicz(1954)}]{lichnerowicz}
\bibinfo{author}{Lichnerowicz, A.}, \bibinfo{year}{1954}.
\newblock \bibinfo{title}{Les Théories relativistes de la gravitation et de l'électromagnétisme}.
\newblock \bibinfo{publisher}{Masson}, \bibinfo{address}{Paris}.
%Type = Book
\bibitem[{Misner et~al.(1973)Misner, Thorne and Wheeler}]{MTW}
\bibinfo{author}{Misner, C.W.}, \bibinfo{author}{Thorne, K.S.}, \bibinfo{author}{Wheeler, J.A.}, \bibinfo{year}{1973}.
\newblock \bibinfo{title}{Gravitation}.
\newblock \bibinfo{publisher}{Princeton University Press}, \bibinfo{address}{Princeton}.
%Type = Article
\bibitem[{Peres(1963)}]{peres}
\bibinfo{author}{Peres, A.}, \bibinfo{year}{1963}.
\newblock \bibinfo{title}{Polynomial expansion of gravitational lagrangian}.
\newblock \bibinfo{journal}{Nuovo Cimento} \bibinfo{volume}{28}, \bibinfo{pages}{865--867}.
%Type = Article
\bibitem[{Rindler(1966)}]{rindler}
\bibinfo{author}{Rindler, W.}, \bibinfo{year}{1966}.
\newblock \bibinfo{title}{Kruskal space and the uniformly accelerated frame}.
\newblock \bibinfo{journal}{Am. J. Phys.} \bibinfo{volume}{34}, \bibinfo{pages}{1974--1978}.
%Type = Book
\bibitem[{Visser(1996)}]{visser}
\bibinfo{author}{Visser, M.}, \bibinfo{year}{1996}.
\newblock \bibinfo{title}{Lorentzian Wormholes: from Einstein to Hawking}.
\newblock \bibinfo{publisher}{AIP Melville}, \bibinfo{address}{New York}.
%Type = Article
\bibitem[{Wang(2023)}]{wang}
\bibinfo{author}{Wang, Z.L.}, \bibinfo{year}{2023}.
\newblock \bibinfo{title}{On a schwarzschild-type defect wormhole}.
\newblock \bibinfo{journal}{arxiv.org/abs/2307.01678} .

\end{thebibliography}



%% For example:

%% \bibitem[Aladro et al.(2015)]{Aladro15} Aladro, R., Martín, S., Riquelme, D., et al. 2015, \aas, 579, A101

%%\end{thebibliography}

\end{document}